\documentclass[preprint2]{aastex62}
\usepackage{natbib}
\usepackage{hyperref}
\usepackage{amsmath}
\usepackage{multirow}
\usepackage{bm}
\usepackage{graphicx}

\begin{document}

\title{New galaxy UV luminosity constraints on warm dark matter from JWST}

\correspondingauthor{Huanyuan Shan}
\email{hyshan@shao.ac.cn}

\author{Bin Liu$^{1}$, Huanyuan Shan$^{1,2,3}$ and Jiajun Zhang}

\affiliation{Shanghai Astronomical Observatory, Chinese Academy of Sciences, Nandan Road 80, Shanghai 200030, China\\}
\affiliation{School of Astronomy and Space Science, University of Chinese Academy of Sciences, Beijing 100049, China\\}
\affiliation{Key Laboratory of Radio Astronomy and Technology, Chinese Academy of Sciences, A20 Datun Road, Chaoyang District, Beijing, 100101, P. R. China\\}

\begin{abstract}
We exploit the recent {\it James Webb Space Telescope} (JWST) determination of galaxy UV luminosity functions over the redshift range $z=9-14.5$ to derive constraints on warm dark matter (WDM) models. The delayed structure formation in WDM universes make high-redshift observations a powerful probe to set limits on the particle mass $m_\mathrm{x}$ of WDM candidates. By integrating these observations with blank-field surveys conducted by the {\it Hubble
Space Telescope} (HST) at redshifts $z=4-8$, we impose constraints on both astrophysical parameters ($\beta$, $\gamma$, $\epsilon_{\mathrm N}$, $M_c$ for a double-power law star formation efficiency, and $\sigma_{M_{\mathrm{UV}}}$ for a Gaussian magnitude-halo mass relation) and the WDM parameter (dark matter particle mass $m_\mathrm{x}$) simultaneously. We find a new limit of $m_\mathrm{x} \geq 3.2$ keV for the mass of thermal relic WDM particles at $95\%$ confidence level. This bound is tighter than the most stringent result derived using HST data before. Future JWST observations could further reduce the observation uncertainties and improve this constraint.  
  
\end{abstract}


\section{Introduction}

Understanding the nature of dark matter remains one of the most important challenges in the field of cosmology. Despite dark matter making up $\sim 85\%$ of matter content in the universe, definitive detection of dark matter remains elusive. While the cold dark matter (CDM) cosmological model is notably consistent with most observational data, small-scale challenges (see e.g. \citealt{Bullock2017}) also prompt exploration of dark matter models beyond standard CDM.

Many alternative candidates such as warm dark matter (WDM), fuzzy dark matter or self-interacting dark matter (\citealt{Bode2001,Hu2000,Spergel2000}) have been proposed to give a better explanation of structure formation at small scales. These scenarios produce a suppression in the matter power spectrum and hold significant implications for various cosmological observations. Constraints on these non-standard dark matter models have been derived from various astrophysical probes, including the Lyman-$\alpha$ forest \citep{Viel2013,Baur2016,Irsic2017,Garzilli2021,Villasenor2023}, $\gamma-\rm{ray}$ bursts \citep{deSouza2013}, strong gravitational lensing 
 \citep{Gilman2019,Gilman2020,Shevchuk2023}, Milky Way satellite galaxies \citep{Kennedy2014,Nadler2021b,Nadler2021a,Newton2021}, neutral hydrogen 21cm signal \citep{Chatterjee2019,Rudakovskyi2020}, cosmic reionization \citep{Barkana2001,Lapi2015,Dayal2017}, and high-redshift galaxy number counts, luminosity function or stellar mass density \citep{Pacucci2013,Schultz2014,Dayal2015,Corasaniti2017,Gandolfi2022,Maio2023,Dayal2024}.

The galaxy UV luminosity function (UVLF) is considered a potent tool for constraining warm dark matter models. In fact, these models yield an exponential suppression on small-scale structures at early cosmic time. The detection of high-redshift galaxies can capture this characteristic and establish stringent limits on WDM particle mass. During the {\it Hubble
Space Telescope} (HST) era, UVLF observations have already reached redshift $z=10$, and tight lower boundaries have been obtained from these results \citep{,Menci2016b,Menci2016a,Menci2017,Rudakovskyi2021}. As anticipated by these studies, observations from the {\it James Webb} Space Telescope (JWST) will further enhance constraint power on the particle mass in WDM models.
The early data releases of JWST have revealed the detection of very distant galaxies up to $z=14-16$. These detections enable us to investigate the galaxy formation and cosmology in the ultra-high-redshift universe. Some intriguing results have emerged (e.g. \citealt{Labbe2023}) and have been extensively discussed (e.g. \citealt{Boylan2023,Wang2023,Parashari2023,Sabti2024}). Meanwhile, early programs has reliably mapped UVLF at redshift $z\geq10$ (e.g. \citealt{Castellano2022,Naidu2022,Finkelstein2023,Casey2023,Adams2023,Bouwens2023,Donnan2023,Harikane2023}), enabling us to constrain the nature of dark matter based on the properties of the first generation galaxies in the universe. Recently, \cite{Donnan2024} combined several major Cycle-1 JWST imaging programs to compile a large and deep sample. Taking advantage of the larger region coverage, the accuracy of UVLF measurement is markedly improved. In this paper, we utilize the new UVLF measurements from JWST as well as the previous results from HST to perform new constraint on WDM models.

This paper is structured as follows. In Sec. \ref{Method}, we describe our UVLF model and data sets. In Sec. \ref{results}, we present the results derived from our analysis, followed by conclusions drawn in Sec. \ref{conclusion}.

Throughout this paper, we assume the fiducial model is a flat $\Lambda$CDM cosmology with $H_0=67.4 \ \rm {km\ s^{-1} \ Mpc^{-1}}$, $\Omega_m=0.315$, $\Omega_b=0.049$, $\sigma_8=0.811$ and $n_s=0.965$.

\section{Methodology}\label{Method}

The UVLF is defined as the number density of galaxies as a function of 
 their UV magnitude. By connecting galaxy properties to dark matter halos, the UVLF can be written as:
 \begin{equation}
\Phi_{\mathrm{UV}}=\frac{\mathrm{d} n_{\mathrm{h}}}{\mathrm{d} M_{\mathrm{h}}} \times \frac{\mathrm{d} M_{\mathrm{h}}}{\mathrm{d} M_{\mathrm{UV}}}
\end{equation}
with the first term $\frac{\mathrm{d} n_{\mathrm{h}}}{\mathrm{d} M_{\mathrm{h}}}$ is the halo mass function, while the subsequent function $\frac{\mathrm{d} M_{\mathrm{h}}}{\mathrm{d} M_{\mathrm{UV}}}$ establishes the connection between halo mass and galaxy luminosity.

To construct the theoretical UVLF, we first need to know the dark matter halo mass function. The halo mass function characterizes the mass distribution of dark matter halos. In the case of CDM, the halo mass function is expressed as:
\begin{equation}
\frac{\mathrm{d} n_{\mathrm{h}}}{\mathrm{d} M_{\mathrm{h}}}=\frac{\bar{\rho}_{\mathrm{m}}}{M_{\mathrm{h}}} \frac{\mathrm{d} \ln \sigma_{M_{\mathrm{h}}}^{-1}}{\mathrm{~d} M_{\mathrm{h}}} f\left(\sigma_{M_{\mathrm{h}}}\right)
\end{equation}
Here $\bar{\rho}_{\mathrm{m}}$ denotes the average comoving matter energy density, and $\sigma_{M_{\mathrm{h}}}$ is the root-mean-square of the density field smoothed over a mass scale $M_h$, typically computed using a top-hat window function and the linear power spectrum. The linear power spectrum can be calculated using publicly available codes 
 $\texttt{CLASS}$\citep{Blas2011} or $\texttt{CAMB}$ \citep{Lewis2000}. The function $f\left(\sigma_{M_{\mathrm{h}}}\right)$ is obtained analytically or measured from simulations. We adopt the Sheth-Tormen form \citep{Sheth2002} in our models.

In WDM scenarios, the halo mass function experiences suppression below a characteristic halo mass. Various studies have suggested that this suppression can be fitted by a parameterized function, as derived from comparisons between CDM and WDM simulations \citep{Dunstan2011,Schneider2012,Lovell2014,Lovell2020}:

\begin{equation}\label{eq:WDM fitting}
n_{\mathrm{WDM}} / n_{\mathrm{CDM}}=\left(1+\left(a M_{\mathrm{hm}} / M_\mathrm{WDM}\right)^b\right)^c
\end{equation}
where $M_\mathrm{WDM}$ is the halo mass of WDM halos, while $a$, $b$ and $c$ are fitting parameters. $M_{\mathrm{hm}}=\frac{4 \pi}{3} \bar\rho_m\left(\frac{\lambda_{\mathrm{hm}}}{2}\right)^3$ denotes the half-model mass where the halo mass function is suppressed by
about a factor of 2. This mass can be quantified using the WDM particle mass $m_\mathrm{x}$ through the half-mode length scale $\lambda_{\mathrm{hm}}$:

\begin{equation}
\lambda_{\mathrm{hm}} \approx 1.015\left(\frac{m_\mathrm{x}}{\mathrm{keV}}\right)^{-1.11}\left(\frac{\Omega_\mathrm{WDM}}{0.25}\right)^{0.11}\left(\frac{h}{0.7}\right)^{1.22}
\end{equation}
with $\Omega_{\mathrm {WDM}}$ representing the WDM background overdensity. In this work, we evaluate the fitting function using the results reported in \cite{Stucker2022}.



The function $\frac{\mathrm{d} M_{\mathrm{h}}}{\mathrm{d} M_{\mathrm{UV}}}$ connects the halo mass with the galaxy luminosity. The UV luminosity of a galaxy is correlated with its star formation rate through \citep{Madau2014}:
\begin{equation}
\operatorname{SFR}=\mathcal{K}_{\mathrm{UV}} \times L_{\mathrm{UV}}
\end{equation}
Assuming a Salpeter initial mass function\citep{Salpeter1955}, the conversion factor is determined to be $\mathcal{K}_{\mathrm{UV}}= 1.15 \times 10^{-28}$. On the other hand, we can also connect star formation rate with halo mass through the baryonic accretion rate:
\begin{equation}
\mathrm{SFR}=f_* \times \dot{M}_{\mathrm{b}}
\end{equation}
Following \citep{Sun2016}, we adopt the baryonic accretion rate as:
\begin{equation}
\dot{M}_{\mathrm{b}} \approx 3 \mathrm{M_{\odot}} / \mathrm{yr}\left(\frac{M_\mathrm{h}}{10^{10} \mathrm{M}_{\odot}}\right)^\delta\left(\frac{1+z}{7}\right)^\eta
\end{equation}
with $\delta=1.127$, $\eta=2.5$ and $M_\mathrm{h}$ representing the halo mass. $f_*$ is the star formation efficiency (SFE), and we employ a double-power law model:
\begin{equation}
f_*=\frac{2\epsilon_{\mathrm{N}}}{\left(\frac{M_{\mathrm{h}}}{M_\mathrm{c}}\right)^{\beta}+\left(\frac{M_{\mathrm{h}}}{M_\mathrm{c}}\right)^\gamma}
\end{equation}
where $\epsilon_{\mathrm{N}}$ is the amplitude of SFE, while $\beta$ and $\gamma$ regulate the slope of SFE at low and high masses, respectively. $M_{\mathrm{c}}$  determines the halos mass at which the SFE peaks.

Combing the above equations, we can get the UVLF as a function of the WDM model parameter as well as the astrophysical parameters. Noting that this function relates the luminosity of a galaxy to a particular halo mass and is considered an average effect, it is necessary to account for the scatter in the magnitude-halo mass relation. Simulations have demonstrated that this scatter is non-negligible \citep{Tacchella2018}, and a halo mass-dependent scatter can even be used to explain the new discoveries of JWST \citep{Sun2023}. Here, we incorporate a Gaussian form stochasticity in the function, as demonstrated in other studies (e.g. \citealt{Sabti2022}):

\begin{equation}
P\left(M_{\mathrm{UV}}\right)=\frac{1}{\sqrt{2 \pi} \sigma_{M_{\mathrm{UV}}}} \exp \left[-\frac{\left(M_{\mathrm{UV}}-\left\langle M_{\mathrm{UV}}\right\rangle\right)^2}{2 \sigma_{M_{\mathrm{UV}}}^2}\right]
\end{equation}
where $\left\langle M_{\mathrm{UV}}\right\rangle$ is the average magnitude calculated from the one-to-one luminosity-halo relation  described above, and $\sigma_{M_{\mathrm{UV}}}$ is the stochasticity parameter, which we treated as a free parameter in our fitting process. Taking stochasticity into consideration, the final luminosity function is:

\begin{equation}
\resizebox{0.98\hsize}{!}{$
\begin{aligned}
\Phi_{\mathrm{UV}}\left(z, M_{\mathrm{UV}}, \boldsymbol{\theta}\right)=&\frac{1}{\Delta M_{\mathrm{UV}}} \int_0^{\infty} \mathrm{d} M_{\mathrm{h}}\left[\frac{\mathrm{d} n_{\mathrm{WDM}}}{\mathrm{d} M_{\mathrm{h}}}\left(z, M_{\mathrm{h}}, \boldsymbol{\theta}\right)\right.\\ 
&\left.\times \int_{M_{\mathrm{UV}, 1}}^{M_{\mathrm{UV}, 2}} \mathrm{~d} M_{\mathrm{UV}}^{\prime} P\left(M_{\mathrm{UV}}^{\prime}, z, M_{\mathrm{h}}, \boldsymbol{\theta}\right) \right]
\end{aligned}
$}
\end{equation}
with $\boldsymbol{\theta}$ representing the parameter set in our model, $\boldsymbol{\theta}=\{\epsilon_\mathrm{N},\beta,\gamma,M_\mathrm{c},\sigma_{M_{\mathrm{UV}}},m_\mathrm{x}\}$.

To estimate these parameters, we perform the analysis with Markov chain Monte-Carlo (MCMC) sampling via Python package \texttt{emcee} \citep{Foreman-Mackey2013}. We use a modified version of the $\texttt{GALLUMI}$ \citep{Sabti2022} likelihood:

\begin{equation}
-2 \ln \mathcal{L}=\sum_{M_{\mathrm{UV}},z}\left(\frac{\Phi_{\text {model }}\left(z, M_{\mathrm{UV}}, \boldsymbol{\theta}\right)-\Phi_{\mathrm{data}}}{\sigma_{\Phi}^{\text {data }}}\right)^2
\end{equation}
where $\Phi_{\text {model}}\left(z, M_{\mathrm{UV}}, \boldsymbol{\theta}\right)$ is the theoretical luminosity function at each magnitude bin and redshift computed above, and $\Phi_{\mathrm{data}}$ and $\sigma_{\Phi}^{\text {data }}$ are the measured UVLF and errors, respectively. 
The priors for our parameters are listed in Table. \ref{tab:Priors of parameters}. It's worth noting that we use $1/m_{\mathrm{x}}$ instead of $m_{\mathrm{x}}$ in the fitting. This choice is made because the lager values of $m_{\mathrm{x}}$ tend to converge to CDM model results, and there are no reliable upper limits.

\begin{table}[h!]
    \centering
    \begin{tabular}{cccc}
    \hline
    Parameter & Prior Value & Units & Prior form \\
    \hline$\epsilon_{\mathrm{N}}$ & $10^{-3}-1.0$ & - & flat log \\
    $\beta$ & $-3.0-0.0$ & - & flat linear \\
    $\gamma$ & $0.0-3.0$ & - & flat linear \\
    $M_{\mathrm{c}}$ & $10^7-10^{15}$ & $\mathrm{M}_{\odot}$ & flat log \\
    $\sigma_{\mathrm{UV}}$ & $10^{-3}-3.0$ & - & flat linear \\ 
    $1 / m_{\mathrm{x}}$ & $0-1$ & $\mathrm{keV}^{-1}$ & flat linear \\
    \hline
    \end{tabular}
    \caption{The parameters and their priors used in our analysis.}
    \label{tab:Priors of parameters}
\end{table}

\begin{figure}[hbp]
     \centering
     \includegraphics[width=0.48\textwidth]{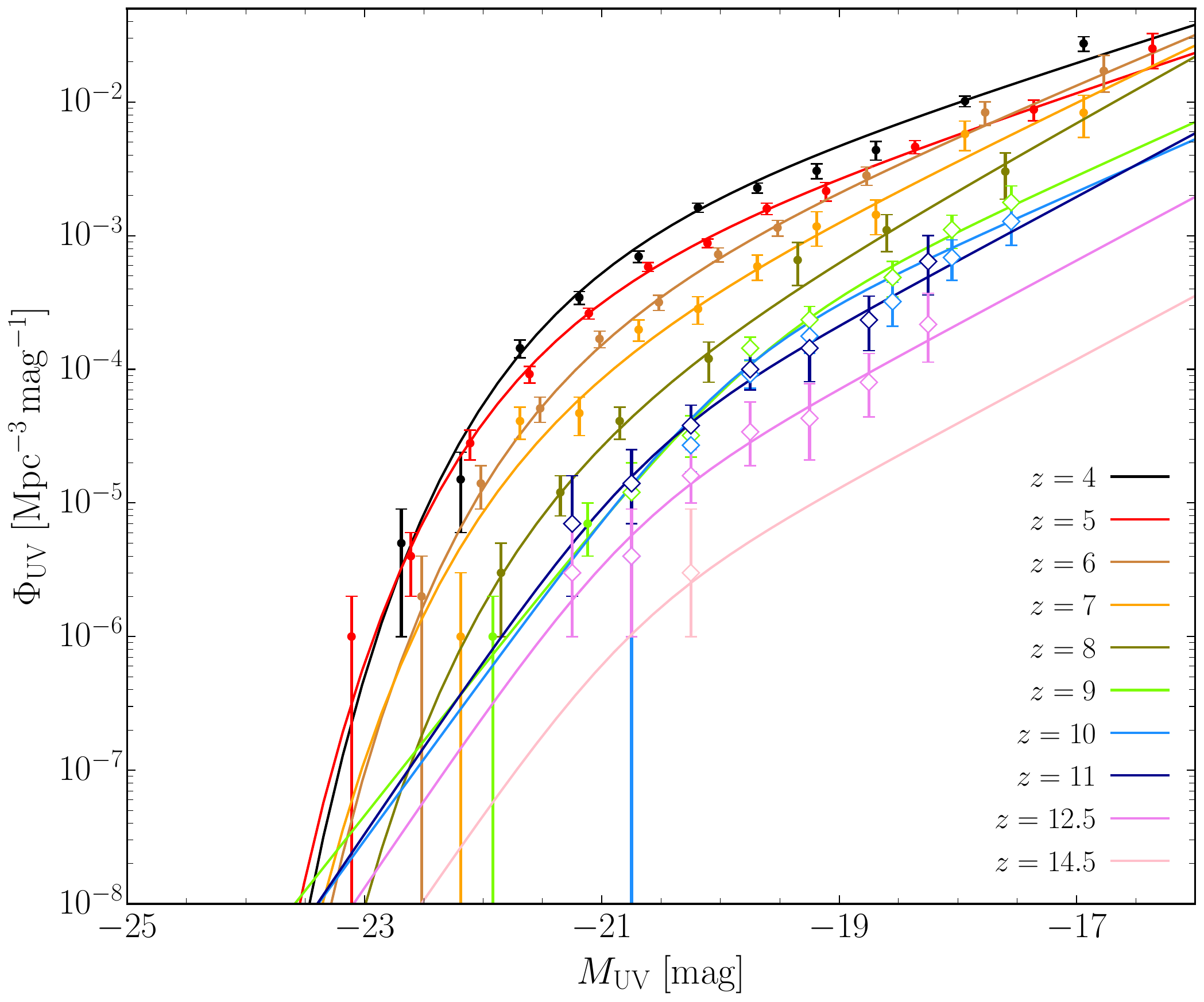}
     \caption{The UVLF data obtained from HST and JWST. HST data are marked with solid circles, while JWST data are marked with open diamonds. Solid lines represent the Schechter or Double-power law fits to these data.}
     \label{Fig:UVLF_data}
\end{figure}

The dataset encompasses both HST and JWST observations spanning the  redshift range $z=4$ to $z=14.5$. JWST has extended the detection of UVLF to higher redshifts. Recently, \cite{Donnan2024} combined several major Cycle-1 JWST imaging programs -- PRIMER (the Public Release IMaging for Extragalactic Research), JADES (the JWST Advanced Deep Extragalactic Survey), and NGDEEP (the Next Generation Deep Extragalactic Exploratory Public survey) -- to produce a new determination. This determination covers 4 separate fields, including the COSMOS and UDS fields (PRIMER), the HUDF-Par2 field (NGDEEP), and the GOODS-South field centred on the HUDF (JADES). It yields a UVLF measurement over an area of $\sim 370$ square arcminutes across the redshift range $z=9-14.5$. Combining these new JWST results with previous HST determinations obtained from the HUDF, the HFF parallel fields, and five CANDELS fields \citep{Bouwens2021} at redshift $z=4-8$, we create a comprehensive sample spanning the widest redshift range for our analysis. The final sample is shown in Figure.\ref{Fig:UVLF_data}. Additionally, We account the effects of dust extinction and apply a correction to the lower redshift and brighter magnitude ends of the data. Assuming the $\rm{IBX}-\beta$ relation, the extinction parameter is then calculated as:

\begin{equation}
\left\langle A_{\mathrm{UV}}\right\rangle=C_0+0.2 \ln (10) \sigma_\beta^2 C_1^2+C_1\langle\beta_{\mathrm{dust}}\rangle
\end{equation}
and we adopt the values of the parameters $C_\mathrm{0}$, $C_\mathrm{1}$, $\sigma_\mathrm{\beta}$ and $\beta_\mathrm{dust}$ following \cite{Sabti2022}.

\section{results}\label{results}
\begin{figure*}[ht]
     \centering
     \includegraphics[trim=0cm 0cm 0cm 0cm, scale=0.4]{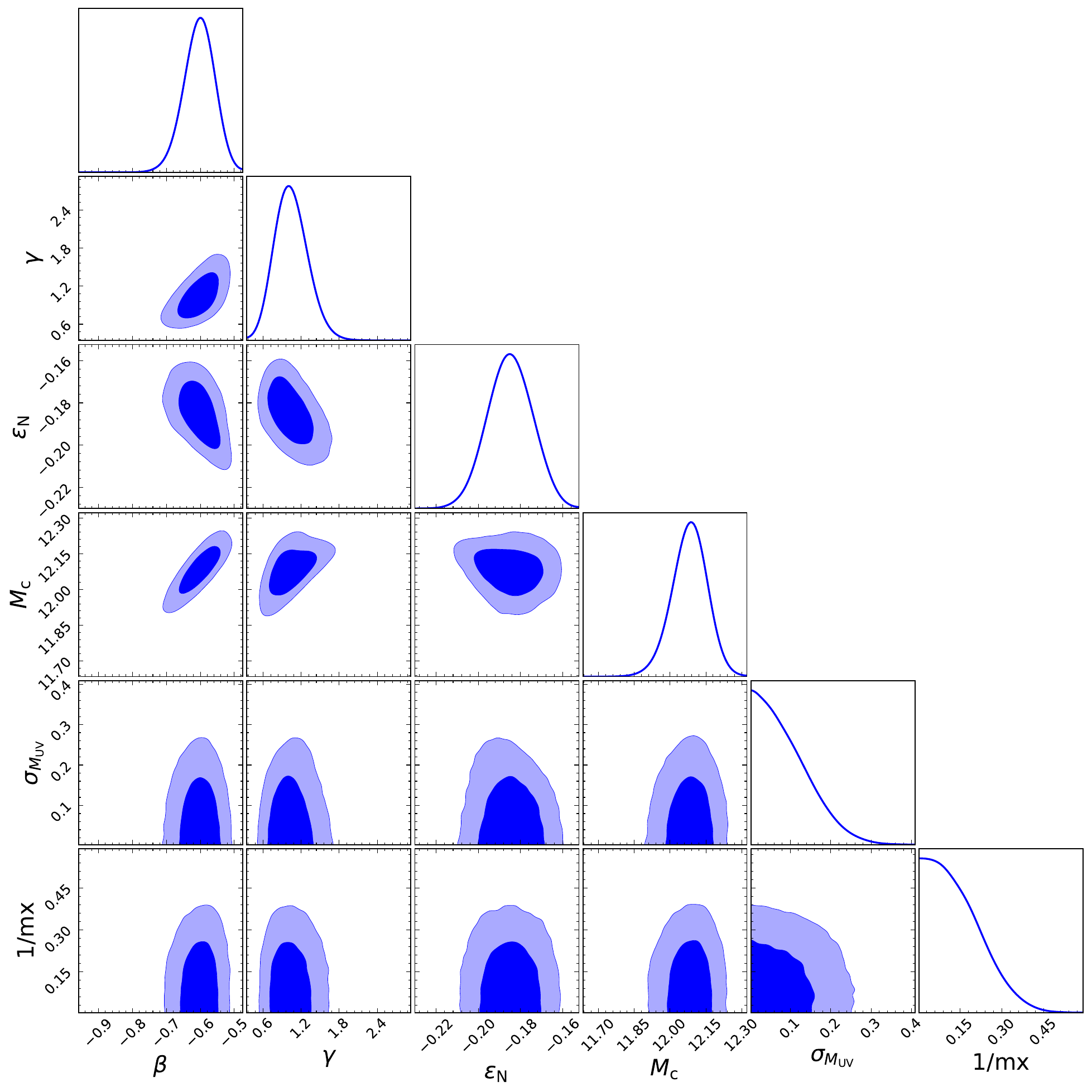}
     \caption{Posterior distributions of the astrophysical parameters and the inverse of dark matter particle mass. The contours show $68\%$
and $95\%$ confidence intervals. The lower bound of WDM particle mass parameter reaches to $\mathrm{m_x} \geq 3.2$ keV at $95\%$ confidence interval.}
     \label{Fig:MCMC_limit}
\end{figure*}

Figure.\ref{Fig:MCMC_limit} presents the MCMC results for all parameters  within our model. The astrophysical parameters are consistent with previous studies (e.g. \citealt{Moster2018,Harikane2022}). Here we seldom pay attention on these parameters and only focus on the dark matter particle mass $m_{\mathrm{x}}$. It is shown that current data cannot rule out WDM models or favor a specific WDM particle mass. Typically, a lower limit is given by UVLF or other probes. The bottom right panel of Figure.\ref{Fig:MCMC_limit} shows the marginalized 1D posteriors of $m_\mathrm{x}$. As illustrated in this figure, the $95\%$ credible limit reaches $m_\mathrm{x} \geq 3.2$ keV. This result suggests that JWST observations can further enhance the constraint power on WDM,  aligning with earlier forecast made by \cite{Rudakovskyi2021} based on simulated JWST results (\citealt{Park2020}). Our lower bound is even tighter than their prediction due to the broader redshift coverage.

\begin{figure}[htbp]
     \centering
     \includegraphics[trim=0cm 0cm 0cm 0cm, scale=0.7]{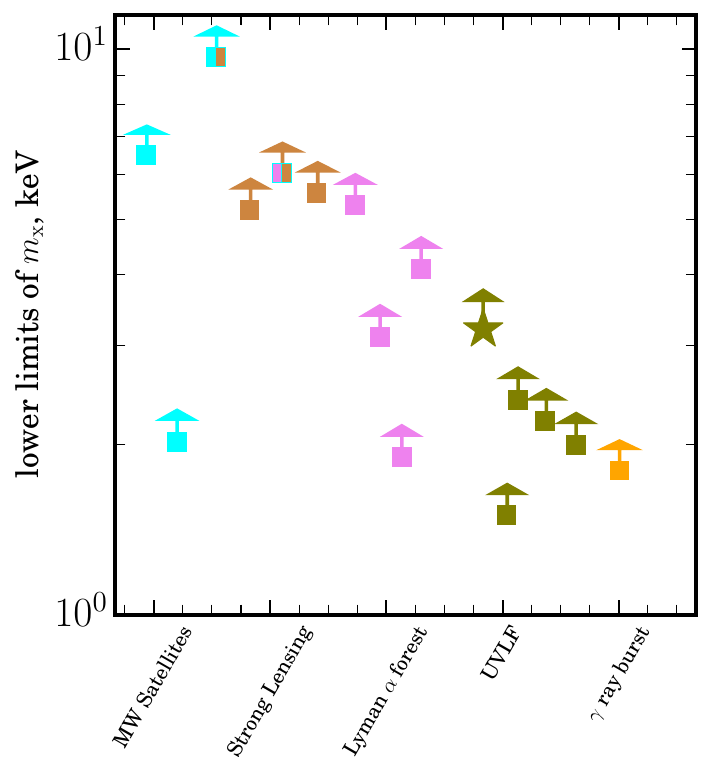}
     \caption{Lower limits of $m_\mathrm{x}$ derived from different probes: Milky Way satellites (cyan, \citealt{Newton2021, Nadler2021a}), strong gravitational lensing (peru, \citealt{Hsueh2020,Gilman2020}), Lyman $\alpha$ forest (violet, \citealt{Baur2016,Irsic2017,Garzilli2021,Villasenor2023}), UVLF (olive, \citealt{Menci2016b,Corasaniti2017,Rudakovskyi2021,Maio2023}), $\gamma$ ray burst (orange, \citealt{deSouza2013}) or joint analysis (mixed color, \citealt{Enzi2021,Nadler2021b}). Star symbol denotes the result derived from this work.}  
     \label{Fig:mx_limits_different_probes}
\end{figure}

Our result updates the tightest constraint derived from the UVLF method. Previous studies have obtained several limits by comparing cumulative galaxy number or fitting analytic models to UVLF observations. The tightest result was obtained by \cite{Menci2016b} using galaxy number counts. Compared to their findings, we derive a tighter constraint with the inclusion of high-redshift JWST data. Figure.\ref{Fig:mx_limits_different_probes} shows our new constraint compared with several tight results derived from other UVLF research \citep{Menci2016b,Corasaniti2017,Rudakovskyi2021,Maio2023}. Furthermore, even tighter limits obtained from the analysis of strongly lensed quasars \citep{Gilman2020,Hsueh2020}, Milky Way Satellites \citep{Nadler2021a}, Lyman-$\alpha$ forest \citep{Baur2016,Irsic2017} or a combination of these probes \citep{Enzi2021,Nadler2021b} are also illustrated. Although the results derived from UVLF are not the most stringent constraints, these results provide comprehensive exploration at high redshift. Meanwhile, robust limits have not been established, conservative estimations of these approaches lead to weaker lower bounds \citep{Garzilli2021,Newton2021}, our constraint can serve as  cross-validation in testing the nature of dark matter.

\section{Conclusion}\label{conclusion}
Current and upcoming JWST surveys can provide improved measurements on the abundance and brightness of galaxies in the early universe. This will enable us to detect the nature of dark matter by making use of the UVLF of galaxies. In this work, we have compiled the most extensive determination of UVLF from JWST, along with previous blank-field surveys conducted by HST, to make a widest UVLF sample spanning from redshift $z=4$ to $z=14.5$. We then perform astrophysical constraints on WDM cosmologies using this sample.

Employing a double-power law astrophysical model, we simultaneously fit the astrophysical and WDM parameters. Based on current observations, we find the $95\%$ credible limit of the lower boundary is $m_\mathrm{x} \geq 3.2$ keV for thermal relic WDM particles. This is the most stringent constraint derived from UVLF and confirms previous forecast that early JWST observations would imply a WDM particle mass of $m_\mathrm{x} > 2.5$ keV. Similar conclusions were also obtained by \cite{Dayal2024} using the properties of stellar mass. By comparing the stellar mass functions, the stellar mass density, and the maximum stellar mass in CDM and WDM cosmologies with observed datasets, they concluded that the 1.5 keV WDM model can be ruled out, and the 3.0 keV WDM model can be distinguished with more high-redshift and low-mass observations. Through the integration of more JWST datasets, our results provide preliminary experiments and corroborate that the ultra-high-redshift detections of JWST will be crucial probes for constraining the mass of WDM particles in the early universe.

The constraint could be further improved in the future by extending the current detection of faint objects to cover a larger area or to deeper magnitude limits. As an example, assuming that the current deepest region ($\sim 30$ AB mag) maintains the same area coverage ($\sim 370$ square arcminutes), the measurement will improve the lower bound to $m_\mathrm{x} \geq 4.3$ keV. Similar constraints could also be obtained with detections extended to redshift $z\sim15$. Future programs could carry out the necessary observations to validate this calculation. Additionally, JWST will not only improve the data quality for bright galaxies, the detection limit will also be extended. With the assistance of cluster lensing magnification, the depth of observations can reach $\sim 31-34$ mag. This will enable us to detect galaxies at even fainter magnitudes as well as higher redshifts beyond $z>15$. Precise and deep observations will yield stronger constraints which would be comparable to the results inferred from lensed quasars, albeit long observational time is required.   

\section{Acknowledgements}

We acknowledge the support by National Key R\&D Program of China No. 2022YFF0503403, No.2022YFF0504300 and the Ministry of Science and Technology of China (grant Nos. 2020SKA0110100). This work is supported by China Manned Space Project (No. CMS-CSST-2021-A01, CMS-CSST-2021-A04, CMS-CSST-2021-A07, CMS-CSST-2023-A03). BL acknowledges the Shanghai Post-doctoral Excellence Program under Grant No. 2022671. HYS acknowledges the support from NSFC of China under grant 11973070, Key Research Program of Frontier Sciences, CAS, Grant No. ZDBS-LY-7013 and Program of Shanghai Academic/Technology Research Leader. JZ acknowledges the support from the China Manned Space Project with no. CMS-CSST-2021-A03. 

\bibliography{refs}

\end{document}